\documentclass[%
 reprint,
 amsmath,amssymb,
prb,
]{revtex4-2}

\usepackage{gensymb}
\usepackage{graphicx}
\usepackage{bm}
\usepackage{hyperref}
\usepackage{dcolumn}

\usepackage{blindtext}
\usepackage{color}

\begin{document}

\preprint{APS/123-QED}

\title{Experimental evidence of isotropic transparency and complete band gap formation for ultrasounds propagating in stealth hyperuniform media}

\author{Ludovic Alha{\"{\i}}tz}
 \affiliation{Sorbonne Université, CNRS, Institut Jean Le Rond d'Alembert, UMR 7190, 4 Place Jussieu, Paris, F-75005, France}
\author{Jean-Marc Conoir}
 \affiliation{Sorbonne Université, CNRS, Institut Jean Le Rond d'Alembert, UMR 7190, 4 Place Jussieu, Paris, F-75005, France}
 \author{Tony Valier-Brasier}\thanks{tony.valier-brasier@sorbonne-universite.fr}
 \affiliation{Sorbonne Université, CNRS, Institut Jean Le Rond d'Alembert, UMR 7190, 4 Place Jussieu, Paris, F-75005, France}
 \email{tony.valier-brasier@sorbonne-universite.fr}

\begin{abstract}
Following on recent experimental characterization of the transport properties of stealth hyperuniform media for electromagnetic and acoustic waves, we report here measurements at ultrasonic frequencies of the multiple scattering of waves by 2D hyperuniform distributions of steel rods immersed in water.
The transparency, for which the effective attenuation of the medium is cancelled, is first evidenced by measuring the transmission of a plane wave propagating in a highly correlated and relatively dense medium. It is shown that a band gap occurs in the vicinity of the first Bragg frequency. The isotropy of both transparency and bang gap are also evidenced for the case of waves generated by a point source in differently ordered and circular shaped distributions. In other words, we thus obtain a representation of the Green's function. Our results demonstrate the huge potential of hyperuniform as well as highly correlated media for the design of functional materials.
\end{abstract}

\maketitle

\section{Introduction}

Wave propagation in complex media is the subject of intense research these past few years in both optical and acoustical communities, with a particular motivation for the control of propagation.
Recent studies proposed to design the disorder itself, by increasing position correlations between scatterers, in order to access unusual properties at different frequency ranges. For example correlations change the scattering mean free path \cite{fraden1990multiple}, may support band gaps \cite{froufe2016role, vynck2021light}, and could also have strong implications in the context of diffusive transport or Anderson localization \cite{froufe2017band, monsarrat2022pseudogap}.

Recently, a new class of correlated media called stealth hyperuniform (SHU) media \cite{gabrielli2002glass, torquato2003local}, have shown a tremendous potential in the enhancement of these properties.  
These are disordered media from a local point of view, but characterized by the cancellation of density fluctuations at a long range scale. It gives them an isotropic behavior as purely random media, but also scattering properties similar to crystalline media. Their robustness against scatterer position defects gives them a considerable advantage over phononic media \cite{froufe2023bandgap}, which partly explains their attractiveness for the design of complex media able to control wave propagation.
The exploration of SHU properties began in optics \cite{leseur2016high, bigourdan2019enhanced, sheremet2020absorption}, and especially two interesting phenomena have been investigated: a transparency regime at low frequencies and the appearance of band gaps at higher frequencies.

As in phononic media, the transparency effect occurs in a broad range of frequencies, resulting from a significant reduction of the scattering.
The robustness of the transparency in the presence of scatterers of finite size as well as polydispersity  has been numerically investigated for elastic and electromagnetic waves in \cite{kim2020multifunctional} and acoustic waves in \cite{rohfritsch2020impact, romero2021wave, kuznetsova2021stealth, cheron2022experimental, cheron2022wave}. 

In this paper, we mainly tackle the band gap phenomena which was thought to appear only in crystalline media. Lot of research has been done to explain the origin of band gaps in disordered media with structural correlations \cite{jin2001photonic,edagawa2008photonic,florescu2009designer,florescu2009complete,imagawa2010photonic,liew2011photonic,froufe2016role, froufe2023bandgap}, as well as the coupling with local resonances as discussed in \cite{vynck2021light, gkantzounis2017hyperuniform, monsarrat2022pseudogap, romero2021wave}. 
Studies in audible acoustics have inspected the wave transmission behavior depending on the medium order and shown that for highly correlated media, band gaps are similar to those appearing in phononic media \cite{kuznetsova2021localized,cheron2022experimental, cheron2022wave}.
Interestingly, numerical characterizations \cite{florescu2009designer, florescu2013optical, gkantzounis2017hyperuniform} and experimental measurements \cite{man2013isotropic, man2013photonic} have shown that these band gaps can be completely isotropic in a wide frequency range contrary to crystalline media.
This characteristic could allow the design of free form wave guides for example \cite{man2013isotropic, gkantzounis2017hyperuniform}.
However, no experimental evidence for the isotropy of transparency and the isotropy of band gap formation has yet been made in acoustics.

We report here experimental ultrasonic measurements on non-resonant SHU media that follow on from other experimental studies carried out in microwaves by the group of G. J. Aubry \cite{aubry2020experimental}, and more recently in acoustics in the case of rigid scatterers disposed in a air filled guide \cite{cheron2022experimental}. Exact calculations of the acoustic field scattered by a large number of cylinders were also carried out by an in-house multiple scattering software called MuScat \cite{rohfritsch2019numerical}.
We aim to evidence experimentally the isotropy of transparency for the propagation of ultrasonic waves in relatively dense distributions of non-resonant cylindrical scatterers immersed in water, as well as the appearance of complete band gap.
For high values of the stealthiness degree $\chi$, the transmission drops are stronger but anisotropic as a consequence of a crystallization of the medium.

This article is organized as follows.
In section \ref{sec:scattering_in_SHU}, we outline the main structural and scattering properties of 2D non-resonant SHU media that have been reported in the recent literature. We also derive the expressions of the transmitted and scattered fields which are used further for the comparison of experimental and numerical quantities.
In section \ref{sec:experimental_transmission}, we show measurements of the transmission through a highly organized rectangular shaped distribution of steel rods immersed in water. The transparency and the appearance of a band gap are observed. These measurements are compared with a numerical MuScat simulation and homogenization models.
In section \ref{sec:isotropy}, we experimentally evidence the isotropic nature of SHU media transport properties. Measurements of the directivity pattern of a spherical field that propagates trough differently organized circular distributions of steel rods are conducted. The isotropy of transparency is evidenced as well as the isotropy of band gaps. For strongly organized SHU media, the bang gap is anisotropic as in the case of crystalline media.

\section{\label{sec:scattering_in_SHU}Scattering by dense hyperuniform distributions of non-resonant scatterers}

\subsection{\label{sec:properties_SHU}Properties of hyperuniform distributions of cylinders}

We consider a square medium of width $L$ composed of $N$ cylinders of radius $a$. The fluid matrix is characterized by its mass density $\rho$, sound speed $c$, and wave number $k=\omega/c$ with $\omega$ the pulsation.
The density of cylinders is defined as $\phi=N \pi a^2/L^2$.

The scattering properties of a distribution of $N$ point scatterers are strongly linked to the structure factor of this distribution which is defined in the reciprocal space as:
\begin{equation}
    S(\bold{q}) = \frac{1}{N}\sum_{i=1}^N \left|\mathrm{e}^{\mathrm{i}\bold{q}\cdot \bold{r}_i } \right|^2.
\end{equation}
$\bold{q}=\bold{k}_\mathrm{scat}-\bold{k}_\mathrm{inc}$ is a vector in the reciprocal space, with $\bold{k}_\mathrm{inc}$ and $\bold{k}_\mathrm{scat}$ the incident and scattered wave vectors.
A ``stealth'' hyperuniform (SHU) point distribution has a structure factor that 
vanishes in the long wavelength limit $|\bold{q}|\rightarrow 0$, except at the origin where $S(\bold{q}=\bold{0})=N$ corresponds to forward scattering $\bold{k}_\mathrm{scat}=\bold{k}_\mathrm{inc}$.

SHU distributions are characterized by a parameter $\chi$, the stealthiness degree, which determines the spatial correlation degree between scatterers. $\chi=0$ leads to a random distribution and its maximal value $\chi=\pi/4$ leads to a perfect crystal \cite{cheron2022experimental}. 
This parameter is defined as the ratio between the number of constrained degrees of freedom in reciprocal space and the total number of spatial degrees of freedoms. One can find a detailed description of the parameter $\chi$ in Ref.\,\cite{uche2004constraints}. 
In this article, SHU media are generated by constraining the structure factor to vanish within a disk of radius $K=\pi\sqrt{4 \chi N +1}/L$ in the reciprocal space, and by following the procedure described by Zhang \textit{et al.} \cite{zhang2015ground1, zhang2015ground2} and Froufe-P{\'e}rez \textit{et al.} \cite{froufe2016role}.

For cylinders of finite size ($a>0$), the cut-off frequency limiting the cancellation of the structure factor is approached as follows \cite{rohfritsch2020impact}:
\begin{equation}
    f_\mathrm{c} = \frac{cK}{4\pi} = \frac{c}{2 a} \sqrt{\frac{\phi \chi}{\pi}}.
    \label{eq:cut-off}
\end{equation}
Each medium can be characterized by a mean distance between scatterers $d=[A/N]^{1/2}$, where $A$ is the area enclosing all the centers of the scatterers. This distance is arbitrary defined as a characteristic distance corresponding to a square lattice, noting that SHU distributions have no periodic lattice.
We also introduce the Bragg vector norm ${q_\mathrm{B}=2\pi/d}$ where the maximum of the structure factor is nearly located \cite{cheron2022experimental}. the Bragg wave number is then $k_\mathrm{B} = q_\mathrm{B}/2$ and the Bragg frequency is 
\begin{equation}
f_\mathrm{B} = \frac{c}{2d}. 
\label{eq:f_Bragg} 
\end{equation}
In the following, this frequency is used as an indication of the vicinity of which the band gap in transmission occurs.
In the case of a rectangular distribution of cylinders, the mean distance is $d=a\sqrt{\pi/\phi}$, giving $f_\mathrm{B}=f_\mathrm{c}/\sqrt{\chi}$. The Bragg frequency $f_\mathrm{B}$ is thus greater than the cut-off frequency of SHU media $f_\mathrm{c}$.

\subsection{Scattering properties of stealth hyperuniform media}

The coherence length in a heterogeneous medium is quantified by the scattering mean free path ${\ell_s(\omega) = 1/2 \alpha_\mathrm{eff}}$, where $\alpha_\mathrm{eff}$ is the effective attenuation of the medium.
Under the Independent Scattering Approximation (ISA), the mean free path is expressed as  ${\ell_s^\mathrm{ISA}(\omega) = [n_0\sigma_\mathrm{s}(\omega)]^{-1}}$, where $n_0$ is the number density of scatterers and $\sigma_\mathrm{s}$ the scattering cross section of a single scatterer found using Mie theory.

In the single scattering regime, the scattered intensity is proportional to the structure factor \cite{leseur2016high}.
If the incident wave vector is such that $\bold{k}_\mathrm{inc}<K/2$, the scattered wave vector $\bold{k}_\mathrm{scat}$ is entirely confined within the disk of radius $K$. 
As a consequence, the scattering is suppressed for all scattering angles at frequencies lower than $f_\mathrm{c}$ given by Eq.\,\eqref{eq:cut-off}. Since the domain where the structure factor vanishes is here a disk, the transparency should be isotropic.
However, in all along this article, we consider relatively dense distributions so that the previous hypothesis might be invalid to describe the acoustic wave transport.
In the multiple scattering regime, Leseur \textit{et al.} have shown that the scattered intensity is not cancelled but only forward scattering occurs for a medium of finite size \cite{leseur2016high}. 
In any case, no losses due to scattering are expected for $f \leq f_\mathrm{c}$, which leads to the cancellation of the effective attenuation $\alpha_\mathrm{eff}$.

\subsection{Transmitted and scattered fields by rectangular SHU media}

A multiple scattering medium can be homogenized by a complex effective wavenumber \cite{derode2006influence}
\begin{equation}
    k_\mathrm{eff}=\frac{\omega}{v_\mathrm{eff}}+\mathrm{i}\alpha_\mathrm{eff},
    \label{eq:keff}
\end{equation}
with $v_\mathrm{eff}$ and $\alpha_\mathrm{eff}$ its effective phase velocity and attenuation.
The coherent pressure field that propagates through the medium is referenced next as total field $p_\mathrm{tot}$. It is the sum of the incident field $p_\mathrm{inc}$ and the coherent scattered field by the medium $p_\mathrm{scat}$.
Assuming that $k_\mathrm{eff}$ is known, the incident field, the coherent total field and the coherent scattered field at a given position $x$ in water are all plane waves that can be expressed in the Fourier domain as follows: 
\begin{align}
    &p_\mathrm{inc}^\mathrm{th}(x,\omega)= \mathrm{e}^{\mathrm{i}k(h+x)},\\
    &p_\mathrm{tot}^\mathrm{th}(x,\omega)= \mathrm{e}^{\mathrm{i}k_\mathrm{eff}h}\mathrm{e}^{\mathrm{i}k x},\\
    &p_\mathrm{scat}^\mathrm{th}(x,\omega)= \mathrm{e}^{\mathrm{i}kx}\left[\mathrm{e}^{\mathrm{i}k_\mathrm{eff}h}-\mathrm{e}^{\mathrm{i}kh}\right].
    \label{eq:fields}
\end{align}
We consider the surrounding medium as a perfect fluid, \textit{i.e.} without attenuation.
In the case of a SHU medium, one can make the hypothesis that $\alpha_\mathrm{eff} \simeq 0$ mm$^{-1}$ for all the frequencies $f<f_\mathrm{c}$. 
In the transparency regime, the coherent total pressure field is then approximated by:
\begin{align}
    &p_\mathrm{tot}^\mathrm{th}(x,\omega)=p_\mathrm{inc}^\mathrm{th}(x,\omega) \mathrm{e}^{\mathrm{i}\omega h(\frac{1}{v_\mathrm{eff}}-\frac{1}{c})}, \;\; \omega<\omega_\mathrm{c}.
    \label{eq:total_field_transp}
\end{align}
The total field is phase shifted from the incident field propagating only in water, but its modulus remains equal to one.
The modulus of the scattered field in the transparency regime is then expressed in the following simplified form:
\begin{align}
    &\left|p_\mathrm{scat}^\mathrm{th}(\omega)\right|= 2 \left|\sin\left(\frac{\omega h}{2}\left[\frac{1}{v_\mathrm{eff}}-\frac{1}{c}\right]\right) \right|, \;\; \omega<\omega_\mathrm{c}. 
    \label{eq:modulus_scattered_field}
\end{align}
If scattering is suppressed, $v_\mathrm{eff}=c$, the medium is therefore non-dispersive. 
However, it should be noted that even in the case where quite significant forward scattering occurs, in dense media for example, we will show that the transmission still remains equal to one as discussed before.

\section{\label{sec:experimental_transmission} Experimental characterization of the transmission}

In this section, we analyze scattering effects exclusively due to the spatial arrangement of cylinders, \textit{i.e.} strong spatial correlations effects, in a wide frequency range from transparency at low frequency up to the first band gap.
The transport properties of this medium are characterized from the transmission of a plane wave through a rectangular shaped distribution of cylinders at normal incidence. 

\subsection{Medium fabrication, ultrasonic set-up and measured quantities}

\begin{figure}[t]%
\centering%
  \includegraphics[width=1\linewidth]{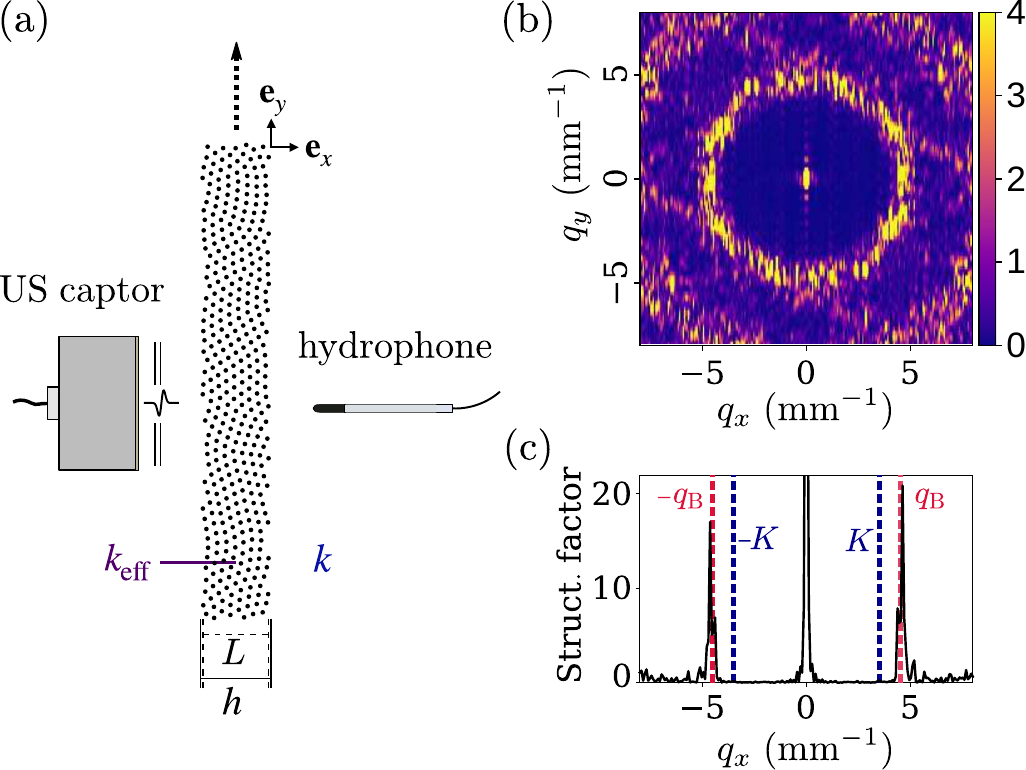}
   \caption{(a) Picture of a part of the SHU distribution used to position steel rods. The density of rods is $\phi=20\%$ and the stealthiness degree is $\chi=0.6$. Burst pulses are emitted in water by an ultrasonic (US) captor and transmitted signals are received by a hydrophone. The distribution is translated along the $y$-axis in order to acquire signals at different positions. (b) Structure factor of the distribution in the reciprocal space. (c) Structure factor along the $q_x$-axis at $q_y=0$ mm${-1}$. The blue dashed lines indicate the vector norm $K$ and the red dotted lines indicate the Bragg vector norm $q_\mathrm{B}$.}
  \label{fig:distrib_config}
\end{figure}

All the media made for the experiments presented in this article are collections of steel rods of same radius $a=0.35$ mm and $200$ mm long. 
The elastic rods are characterized by their density $\rho_\mathrm{p}=8.0$ $\mu$g/mm$^3$, longitudinal wave speed $c^L_\mathrm{p}=5.76$ mm/$\mu$s, and transverse speed $c^T_\mathrm{p}=3.06$ mm/$\mu$s. They are immersed in water of density $\rho=1.0$ $\mu$g/mm$^3$ and sound speed $c=1.49$ mm/$\mu$s. It is a system similar to that used in the characterization of acoustic attenuation in correlated media described in \cite{derode2006influence} for example.
These rods are non resonant at the frequencies of interest ($0.1-0.8$ MHz), hence they are naturally considered as rigid. 
Holes of diameter $0.8$ mm are perforated through thin PMMA plates by following a desired distribution pattern using a laser cutting machine. The rods are then inserted between identically perforated plates that are fixed on a support, which ensures that all the rods are parallel to each other. The support is then immersed in a water tank. 

The transmission experiment is sketched in Fig.\,\ref{fig:distrib_config}(a).  
Short ultrasonic pulses are emitted by a piezoelectric transducer (Olympus, diameter of $25,4$ mm) of $500$ kHz central frequency and with a bandwidth between $100$ kHz and $900$ kHz. 
The dimensions of the transducer are large enough regarding the wavelengths of interest so as to generate a plane wavefront along the acoustic axis.
The ultrasonic signals are recorded by a hydrophone (Teledyne TC4038, global diameter of $4$ mm and with a sensitive area of approximately $3$ mm$^2$) having a flat receiving response in the frequency range $[10-800]$ kHz. The axis of the hydrophone coincides with the axis of the transducer. The hydrophone is coupled to a preamplifier (RESON EC6081) to increase the amplitude of the measured signals compared to noise.
The support is then translated with a motorized translation stage to acquire signals at different lateral positions along the $y$-direction with a spatial step of $1$ mm.

The acquired signals are multiplied by a Tukey window in order to ignore signals reflected between the distribution of steel rods and the surface of the captors or by the edges of the tank. 
The total field can be assessed experimentally from the ratio between the Fourier transform of the average signal propagating through the medium and the spectrum of a reference signal that propagating only in water.
The total and scattered pressure fields are experimentally evaluated by:
\begin{align}
&p_\mathrm{tot} = \frac{\tilde{S}_\mathrm{avg}}{|\tilde{S}_\mathrm{ref}|}, \label{eq:ptot_exp}\\
&p_\mathrm{scat} = \frac{\tilde{S}_\mathrm{avg}- \tilde{S}_\mathrm{ref}}{|\tilde{S}_\mathrm{ref}|},\label{eq:pdiff_exp}
\end{align}
where $\tilde{S}_\mathrm{avg}$ is the Fourier transform of the spatial average of signals which propagate through the distribution, and $\tilde{S}_\mathrm{ref}$ is the one of the reference signal. Note that the transmission coefficient of the medium is therefore obtained from the modulus of the total field.
The effective velocity and attenuation of the medium are determined by the phase difference and the amplitude ratio of the measured acoustic signals as follows:
\begin{align}
    &v_\mathrm{eff} = \frac{\omega h}{kh+\mathrm{arg}(\tilde{S}_\mathrm{avg})-\mathrm{arg}(\tilde{S}_\mathrm{ref})},\label{eq:Veff}\\
    &\alpha_\mathrm{eff} = -\frac{1}{h} \ln\left(\frac{|\tilde{S}_\mathrm{avg}|}{|\tilde{S}_\mathrm{ref}|}\right).\label{eq:Aeff}
\end{align}
 The distance $h=L+2a$ is the propagation distance inside the medium.

\subsection{Transmission through a highly correlated SHU distribution} 

\begin{figure*}[t]%
  \centering%
  \includegraphics[width=1\linewidth]{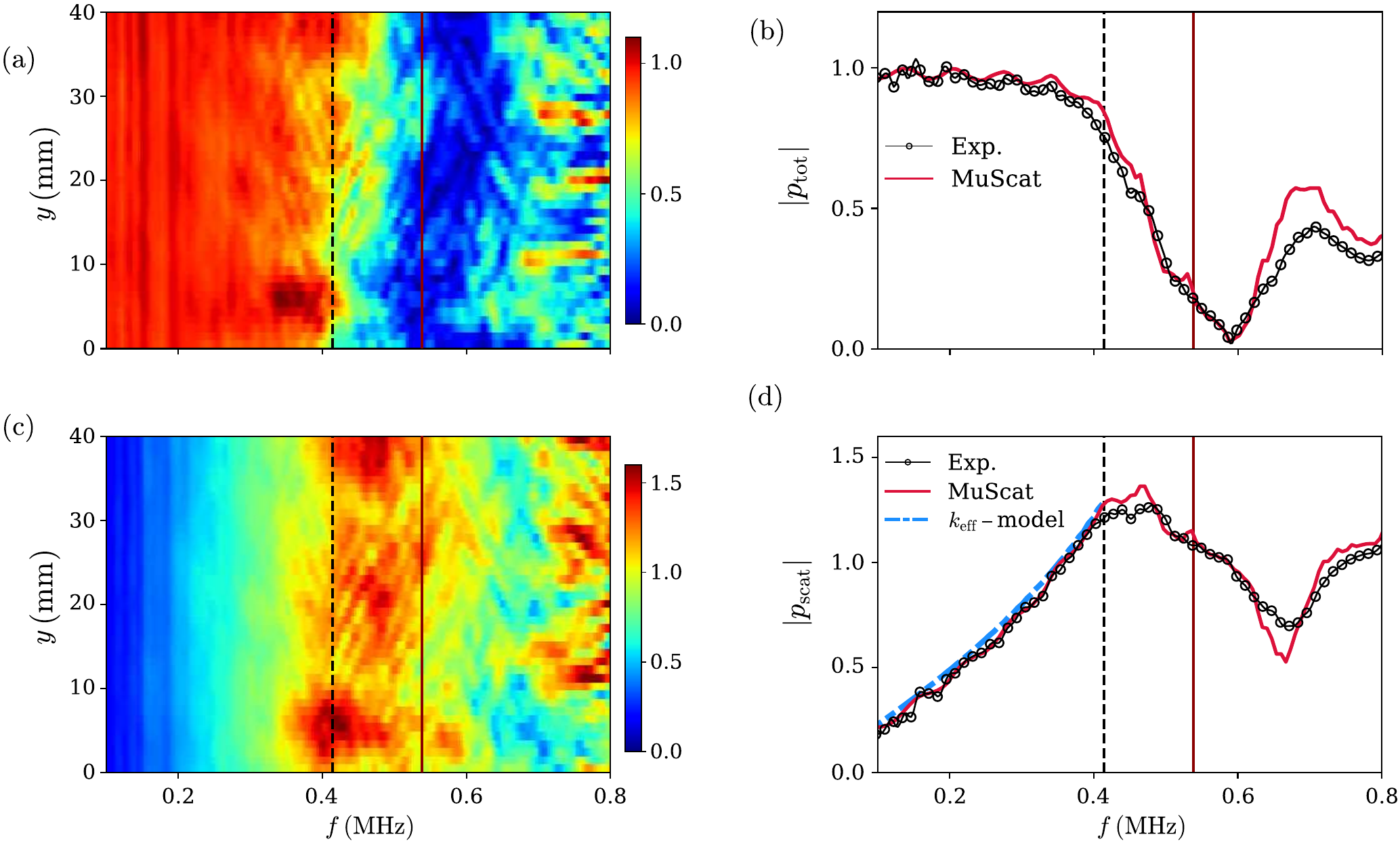}
   \caption{Left: experimental maps of the modulus of (a) total and (c) scattered pressure fields as a function of the frequency and for several lateral positions along the SHU distribution of steel rods, with a spatial step of $1$ mm. Right: mean values of the modulus of (b) total field and (d) scattered field. The vertical black dashed lines indicate the cut-off frequency of the SHU medium $f_\mathrm{c}=0.41$ MHz and the vertical red lines indicate the Bragg frequency $f_\mathrm{B}=0.54$ MHz. Experimental frequency resolution is $3$ kHz.}
  \label{fig:transmi_rect}%
\end{figure*}
 
Numerical and experimental acoustic studies of the influence of $\chi$ on the transmission through 2D SHU media have been recently reported in \cite{rohfritsch2020impact,cheron2022experimental,cheron2022wave}. In particular, it was shown that the transmission remains equal to one for $f \leq f_\mathrm{c}$. After $f_\mathrm{c}$, the amplitude of the coherent total field decreases more and more rapidly and significantly as the parameter $\chi$ increases. It is attributed to a band gap similar to that found in crystalline media \cite{cheron2022experimental}.
We analyze here the transmission through a relatively dense medium of density $\phi=20\%$, with a high value of stealthiness degree $\chi=0.6$. In addition, we are also interested in the dispersion of waves propagating in such a SHU medium in the transparency regime. We thus analyze the evolution of the scattered field and the effective parameters of the medium.

We generate first a square SHU distribution of width $H=150$ mm, composed of $N_t=11691$ scatterers. 
It is then truncated to create a medium of width $L=10$ mm, length $H$, and composed of $N=631$ scatterers.
The cut-off frequency of the distribution is $f_\mathrm{c} = 0.41$ MHz.
The characteristic distance is ${d=[HL/N]^{1/2}=1.38}$ mm, and the corresponding first Bragg frequency is ${f_\mathrm{B}=c /2d=0.54}$ MHz. 
The structure factor of this distribution is shown in Fig.\,\ref{fig:distrib_config}(b) and its evolution along the $q_x$-axis at $q_y=0$ mm$^{-1}$ is plotted in Fig.\,\ref{fig:distrib_config}(c). 
It vanishes within a disk of radius $K=\pi \sqrt{4\chi N_t +1}/H$ and it is maximal on a ring located in the vicinity of the Bragg vector of norm $q_\mathrm{B}=2\pi/d$. 

In the following, the experimental quantities are compared to those obtained with a MuScat simulation performed on the same distribution and also to a Fikioris and Waterman (FW) based model for the determination of $k_\mathrm{eff}$ \cite{fikioris1964multiple,norris2011multiple,rohfritsch2020influence,rohfritsch2021propagation}. This model applies on the coherent wave propagating through a correlated medium by integrating the pair correlation function $g_2(r)$ of the distribution. The calculation of $g_2(r)$ is performed here on the generated square distribution of width $H$. This model is also applied in the case of a purely random medium, considering the Hole Correction ($g_2(r)=0$ for $r<2a$, $g_2(r)=1$ otherwise). In this case, a minimal correlation is however still introduced because the density is quite high and the centers are separated by a minimal distance $2 a$ at least.

Fig.\,\ref{fig:transmi_rect}(a) and (c) show that the amplitude of the measured total and scattered fields is constant along the lateral $y$-direction, almost up to the cut-off frequency $f_\mathrm{c}$. It corresponds exactly to the frequency band for which the structure factor vanishes for $|\bold{q}|\leq K$, identified by the black dotted line at $f=f_\mathrm{c}$. Fig.\,\ref{fig:transmi_rect}(b) and (d) respectively present the spatial averages of the total pressure given by Eq.\,\eqref{eq:ptot_exp} and scattered pressure given by Eq.\,\eqref{eq:pdiff_exp}. First, the MuScat simulation results are in remarkable quantitative agreement with the measurements.
The measured amplitude of the total field for $f$ close to $f_\mathrm{c}$ is one as expected. 
A phase transition happens at $f=f_\mathrm{c}$, as for the structure factor.
We observe then a drastic drop in the amplitude of the total field starting from $f=f_\mathrm{c}$ with a minimal amplitude located near the Bragg frequency $f_\mathrm{B}$. Then, it increases for higher frequencies which is the signature of a band gap. It is worth noting that the location and width of the band gap are approximately the same at each lateral position $y$. It would imply that highly correlated SHU media could have translation invariant properties. Moreover, the scattered field amplitude increases in the transparency band, which is well predicted by Eq.\,\eqref{eq:modulus_scattered_field} and FW model. It leads then to dispersion effects which are analyzed next.


\subsection{Effective parameters of the hyperuniform distribution}

\begin{figure}[t]%
  \centering%
  \includegraphics[width=\linewidth]{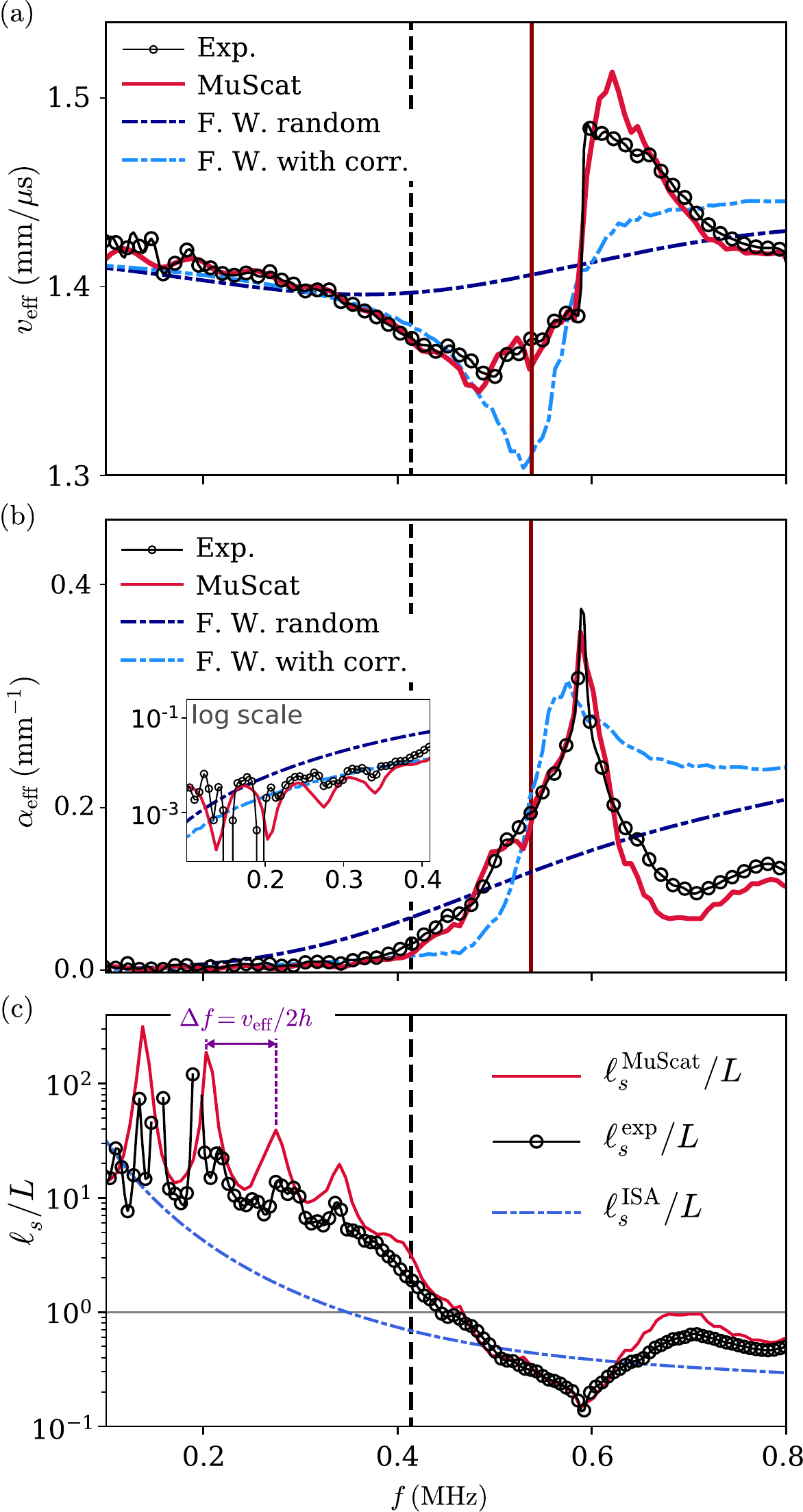}
   \caption{(a) Effective phase velocity and (b) effective attenuation of the SHU medium. The vertical black dashed lines indicate the cut-off frequency $f_\mathrm{c}$ and the vertical red lines indicate the Bragg frequency $f_\mathrm{B}$. (c) Ratio between scattering mean free path $\ell_s$ and medium width $L$. The scattering mean free path determined experimentally and from MuScat is $\ell_s=1/2\alpha_\mathrm{eff}$. The scattering mean free path given by ISA is $\ell_s^\mathrm{ISA}=1/n_0 \sigma_\mathrm{s}$.}
  \label{fig:params}
\end{figure}

The effective phase velocity $v_\mathrm{eff}$ and attenuation $\alpha_\mathrm{eff}$ of the SHU medium are plotted in Fig.\,\ref{fig:params}(a) and (b).
The effective attenuation is quasi null in the transparency regime and reaches a peak in the band gap.
The effective phase velocity slightly decreases in the transparency regime. It means that the medium itself is not 'transparent' as dispersion is observed, but the transmission is still one as predicted by Eq.\,\eqref{eq:total_field_transp}. For $f>f_\mathrm{c}$, it presents a much stronger dispersion with a maximal slope at the frequency corresponding to maximal attenuation.
For both parameters, if only the Hole Correction is applied, the FW model is very far from MuScat simulation and measurements. In the other hand, if the pair correlation function $g_2$ of the distribution is integrated, the FW model captures the evolution of these parameters quite well up to the maximal slope in the velocity and the maximum of attenuation. It highlights strong scattering effects due to short-range correlations. However after this frequency, this model is far from the MuScat simulation and the measurement. This can be explained by the fact that the FW model was initially developed for fully random media. For SHU media, the FW model must be modified to include long-range correlations. This is done here by introducing an integral depending on the pair correlation function. But this is not strictly speaking a completely rigorous approach for SHU media. We may even be surprised that the results are so good. Finally, the most important point is that it shows the effect of correlations.

Fig.\,\ref{fig:params}(c) shows that for $f \leq f_\mathrm{c}$, the mean free path $\ell_s$ is much larger than the distribution width $L$, and also significantly larger than the one predicted by the ISA model which does not take correlations into account. This is hence a clearer evidence of transparency.
We found here that transparency occurs even at a relatively high density ($20 \;\%$), as predicted by Leseur \textit{et al.} \cite{leseur2016high}, and observed in microwave experiments by Aubry \textit{et al.} \cite{aubry2020experimental}. 
The oscillations at low frequencies for $f<f_\mathrm{c}$ can be assumed to be Fabry-Perot interferences due to the finite size of the distribution. The frequency difference between two peaks is $\Delta f = v_\mathrm{eff}/2h=0.067$ MHz. In the other hand, the model considers an infinite homogenized medium. These effects will be investigated in another work.
For the frequencies in the vicinity of $f_\mathrm{B}$, $\ell_s<\ell_s^\mathrm{ISA}$ and both are shorter than $L$ due to strong multiple scattering effects. It generates a lot of destructive interferences at the origin of the observed band gap in transmission.
The band gap effect (periodicity characteristic) is here seen as a drop of the scattering mean free path $\ell_s$ (vision of random medium). In SHU media, we also observe a strong attenuation with dispersion, while periodic media are characterized by a strong attenuation without dispersion.


\section{\label{sec:isotropy}Isotropy behavior of SHU media}

In this part, we aim to show experimentally the isotropic nature of SHU media in the whole frequency range depending on their degree of order. For that purpose, we analyze the angular dependency of the propagation of ultrasonic waves generated by a point source inserted in differently ordered circular hyperuniform distributions of steel rods. 
From another perspective, the experiments presented next allow us to characterize the Green's function of each distribution by measuring the directivity pattern of the radiated field outside.

\subsubsection{Distribution characteristics and experimental procedure}

\begin{figure}[t]%
  \centering%
  \includegraphics[width=1\linewidth]{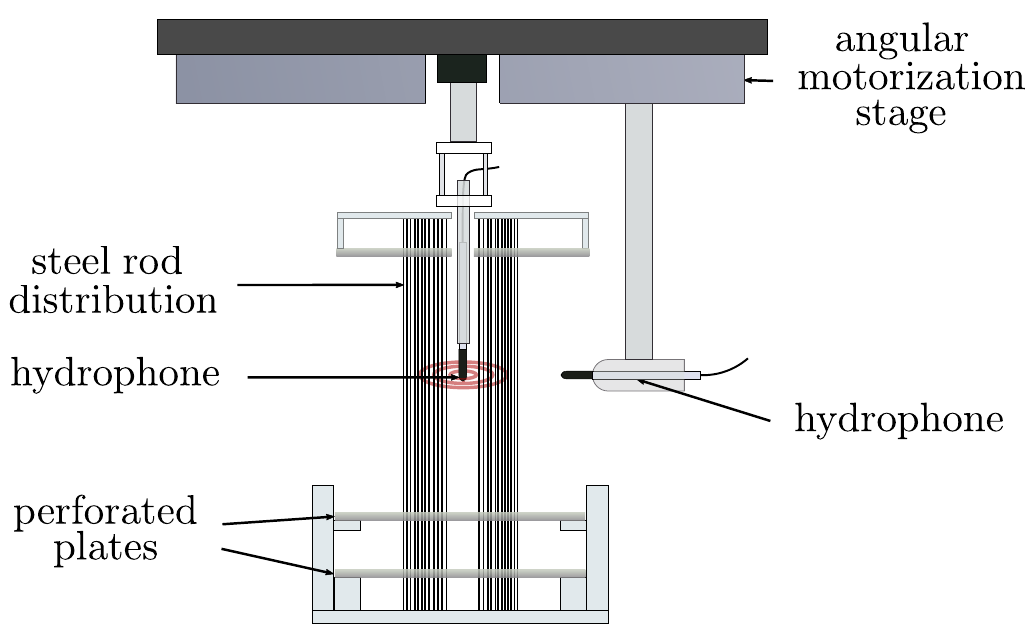}
   \caption{Experimental ultrasonic set up for directivity measurements. A circular medium composed of steel rods inserted between several perforated PMMA plates is deposited in a water tank. A first hydrophone is inserted at the center of the medium and generated an omnidirectional field which propagates through the medium. A second hydrophone is attached to a motorized translation stage and acquires signals at different angular positions.}
  \label{fig:set_up_circ}%
\end{figure}

\begin{figure*}[t]%
  \centering%
  \includegraphics[width=1\linewidth]{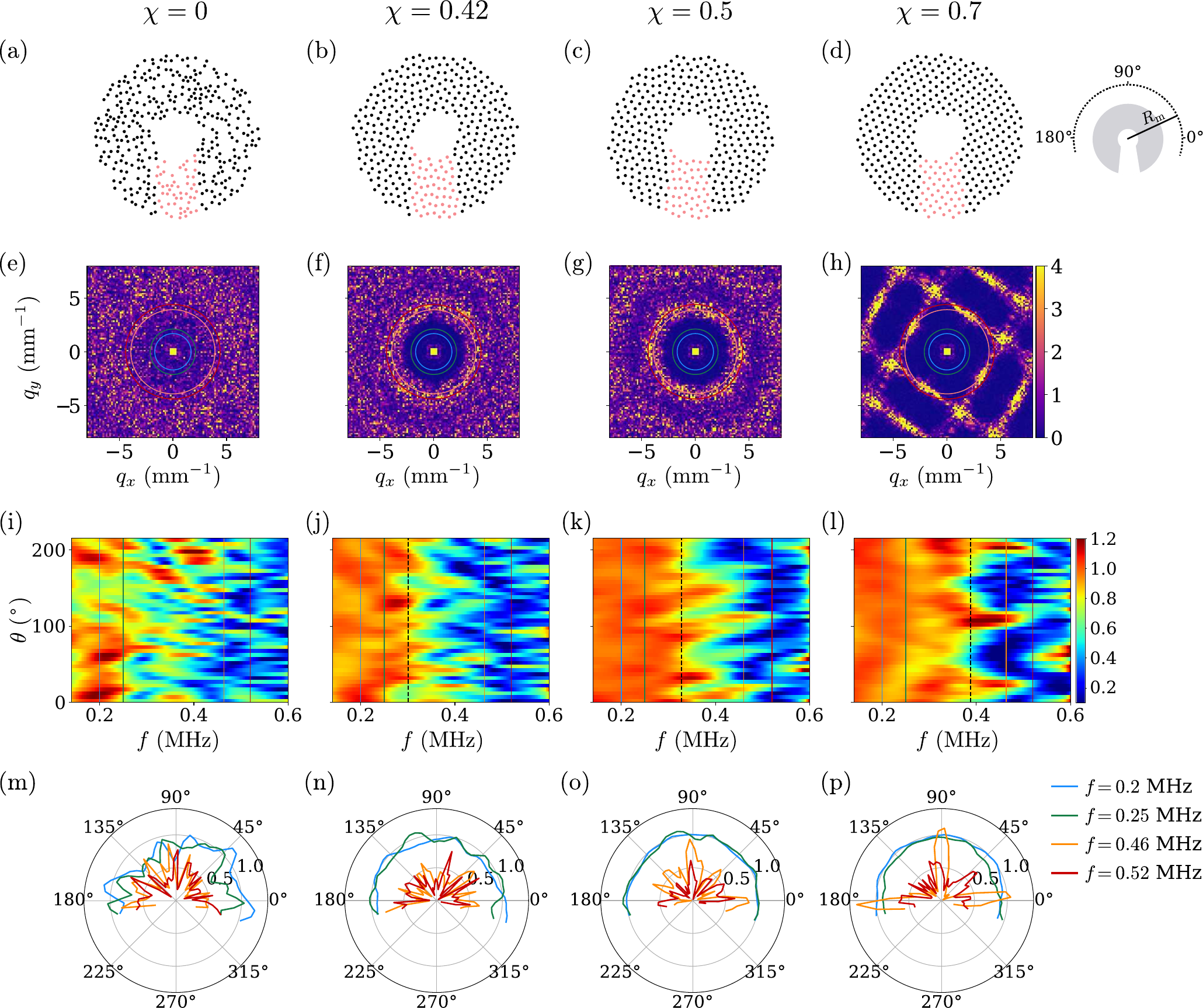}
   \caption{(a)-(d) Distribution patterns with different stealthiness degrees: $\chi=0$, $\chi=0.42$, $\chi=0.5$ and $\chi=0.7$. The red dots are removed to better position the emitting hydrophone at the center of each distribution. (e)-(h) Structure factors of each distribution. The image color is truncated for the sake of visibility. The four circles of different color correspond to different frequencies used for the directivity plots of the total field modulus in (m)-(p): blue $f=0.2$ MHz, green $f=0.25$ MHz, orange $f=0.46$ MHz (Bragg frequency), and red $f=0.52$ MHz. (i)-(l) Maps showing the modulus of the total field at $r=R_\mathrm{m}=28$ mm for the different measurement angles and frequencies. The measurements were performed on an angular range of $215 \degree$ with a step of $5 \degree$ and the frequency resolution is $3$ kHz.}
  \label{fig:transmi_circ}%
\end{figure*}

\begin{figure}[t]%
  \centering%
  \includegraphics[width=\linewidth]{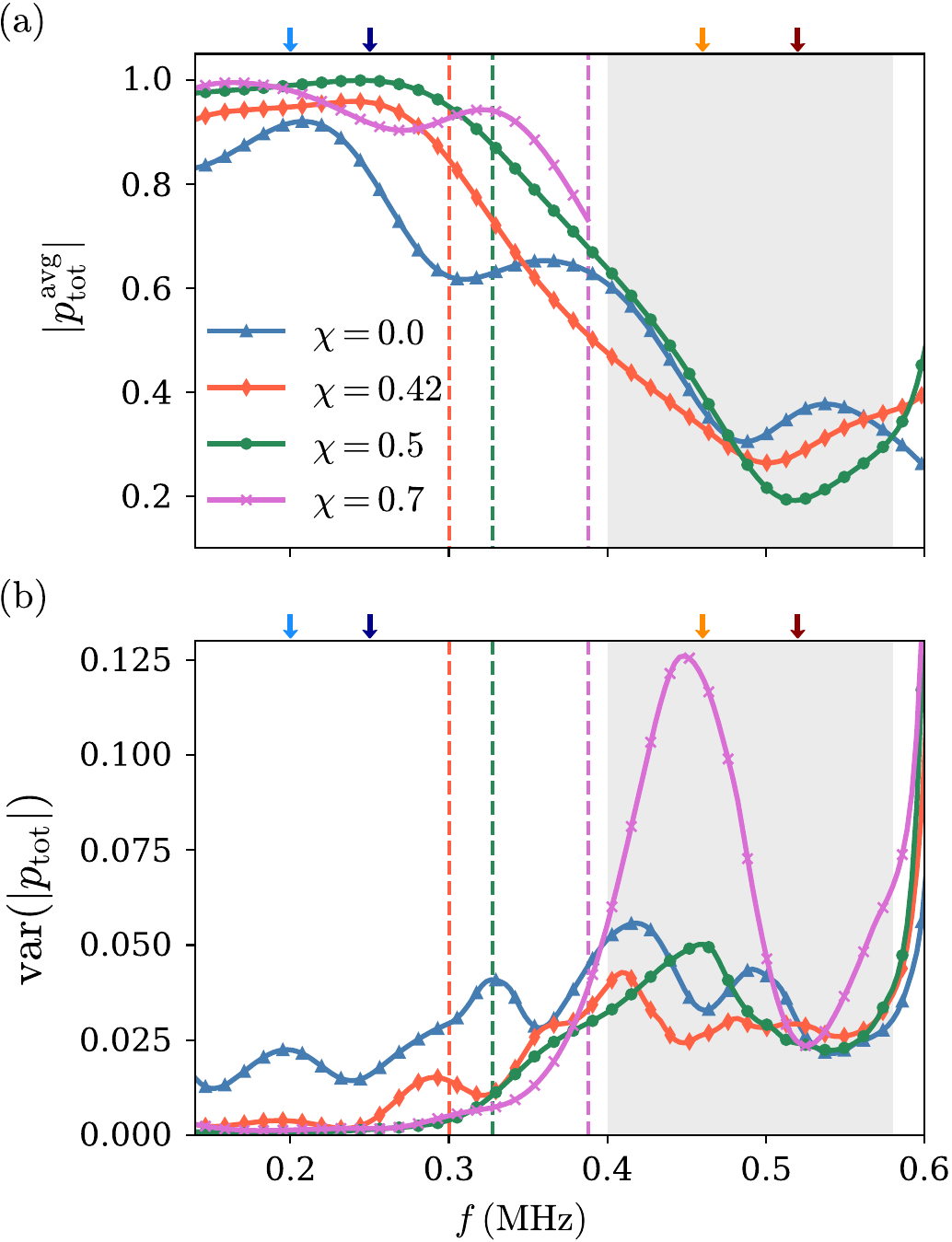}
   \caption{(a) Absolute value of the average total field $p_\mathrm{tot}^\mathrm{avg}$ at $r=R_\mathrm{m}$ for the four distributions studied. The dashed vertical lines show the different cut-off frequencies of each distribution (the line colors refers to the corresponding curves).
   The curve for $\chi=0.7$ is displayed for $f<f_\mathrm{c}$ because the total field has an anisotropic shape for higher frequencies. The gray area covers the frequency band $[0.40-0.58]$ MHz. (b) Variance of the total field modulus about the different angles. Arrows on the top of each figure indicate the frequencies used for the plots in Fig.\,\ref{fig:transmi_circ}(m-p).}
  \label{fig:mean_var_circ}%
\end{figure}

We consider circular-shaped distributions of diameter $R_\mathrm{distrib} = 18$ mm and density $\phi=15\%$ that are displayed in Fig.\,\ref{fig:transmi_circ}(a-d). A random distribution and three distributions of stealthiness degrees $\chi=0.42$, $0.5$ and $0.7$ are generated and used for sample fabrication. 
In order to create an omnidirectional source as incident wavefield, we use a hydrophone (Teledyne TC4038, sensitive area of $3$ mm$^2$) in emission placed at the center of the distribution. This kind of piezoelectric sensor radiates a spherical field with a linear increasing response as the frequency increases.
The center of the distribution is left empty of rods to insert the source. The inner hole has a diameter $R_\mathrm{in} = 5$ mm. The characteristic distance of each distribution is $d=[\pi (R_\mathrm{distrib}^2 - R_\mathrm{in}^2)/N]^{1/2}=1.60$ mm and the corresponding Bragg frequency is $f_\mathrm{B}=c /2d=0.46$ MHz.

The emitting hydrophone is inserted in a rigid pipe with the emitting part being outside and vertically positioned so as to generate waves from the mid-height of the rods as sketched in Fig.\,\ref{fig:set_up_circ}.
The rods are inserted between circular perforated PMMA plates. Two are fixed on a support to maintain the lower part of the rods. A third perforated plate is attached to an other plate placed on the top of the rods to ensure the rod parallelism.
This configuration allows free access all around the distribution and avoids scattering by a potential support bar.
A path without rods is created to put in place the distribution with more precision around the emitting hydrophone, and so as to remove the distribution without moving the hydrophone. The reference measurement is then performed for the exact same source position.
We measure the acoustic pressure field outside the distribution at a distance $R_\mathrm{m}=28$ mm from the center of the distribution with a second identical hydrophone attached to an angular motorization stage. The  emitting hydrophone points towards the axis of rotation of the motor.
The total field is measured over a portion of circle avoiding positions near the rodless path. This portion corresponds to an angular range of $215 \degree$, with an angular step of $5 \degree$.

\subsubsection{Influence of the stealthiness degree on the transport properties} 

We present here results on the propagation of spherical ultrasonic waves in SHU distributions of different stealthiness degrees $\chi$. The structure factors of the four distributions are shown in Fig.\,\ref{fig:transmi_circ}(e-h).
For the random distribution, the non-overlapping of rods naturally results into a minimal distance $2a$ between two rod centers. At high density, short-range correlations naturally appear, that is why the structure factor decreases for small vectors $\bold{q}$. There is however no clear pattern observed in it. In the other hand, SHU media present a structure factor that vanishes on a disk (except at $|\bold{q}|=0$ mm$^{-1}$).
For $\chi=0.42$ and $\chi=0.5$, the structure factor has an isotropic shape and presents rings of amplitude maxima, corresponding to frequencies at which the medium strongly scatters. 
For the medium with $\chi=0.7$, a quasi-crystalline hexagonal lattice state appears.

The transmission properties of multiple scattering media are often related to the shape of their structure factor \cite{leseur2016high}. The moduli of the transmitted fields at $r=R_\mathrm{m}$ are shown in Fig.\,\ref{fig:transmi_circ}(i-l) and the directivity patterns are plotted for several frequencies in Fig.\,\ref{fig:transmi_circ}(m-p)
In the case of the random distribution ($\chi=0$), the amplitude of the field varies significantly around the medium in the entire frequency range.
In the case of all SHU media, we systematically observe a frequency band for which the transmission is quasi one for $f \leq f_\mathrm{c}$. This is the same behavior than for the transmission through the rectangular distribution presented in sec. \ref{sec:experimental_transmission}, with here an omnidirectional incidence. Our results confirm that transparency is isotropic.

As $\chi$ increases, the rings of amplitude maxima in the structure factor have a higher amplitude. As a  result, a drop in amplitude is observed for $f>f_\mathrm{c}$ due to important scattering. 
The transmission minima are here located at higher frequencies than $f_\mathrm{B}$, which suggests that important destructive interference occurs for distances smaller than the characteristic distance of the square lattice $d$.
Importantly, we observe that for $\chi=0.42$ and $\chi=0.5$, the amplitude evenly falls in each direction. It evidences the isotropic nature of the created band gaps.
For $\chi=0.7$, the distribution is quasi-periodic and its structure factor is therefore anisotropic. The scattering occurs then in preferential directions, so the created bang gap is anisotropic too. For example, we observe at $f=0.46$ MHz three of the four expected lobs of maximal amplitude. These are located exactly at the angles where the structure factor is null. On the other hand, the transmission is zero where the structure factor is maximal.

We analyze now statistical quantities to characterize the transport properties of the different media.
The modulus of the average total field over the angles $p_\mathrm{tot}^\mathrm{avg}$ is plotted in Fig.\,\ref{fig:mean_var_circ}(a) for the different values of $\chi$ as a function of the frequency.
For $\chi=0.7$, the plot is made for $f<f_\mathrm{c}$ since the directivity is anisotropic for $f>f_\mathrm{c}$. In this last case, an average after $f_\mathrm{c}$ would therefore be irrelevant for our purpose. 
The modulus of the average total field fluctuates in the random case, while it is quasi one until the cut-off frequency $f_\mathrm{c}$ in the case of SHU distributions. Then, for $f>f_\mathrm{c}$, the amplitude decreases less rapidly but more significantly in the band gap as $\chi$ increases.
In order to characterize the frequency width of isotropy, the variance of the total field modulus about the measured angles is shown in Fig.\,\ref{fig:mean_var_circ}(b). For all SHU distributions, it is quasi null before the respective cut-off frequencies $f_\mathrm{c}$. At higher frequencies, in the frequency band (arbitrary chosen) indicated by a gray background, the variance is low and quasi-constant for all $\chi<0.7$. For $\chi=0.7$, it presents important variations with a peak of high value, which shows that an isotropic behavior occurs in this case in a very limited band. These results suggest then that the Green's function is isotropic in the entire frequency range when $\chi$ is of moderate value, \textit{i.e.} before the crystallization of the medium.

\section{Conclusions}

We have analyzed the transport properties of ultrasounds propagating in stealth hyperuniform distributions of non resonant rods immersed in water.

We conducted a first experiment to demonstrate transparency and band gap formation in a highly correlated SHU medium. In addition, MuScat simulation results are in remarkable quantitative agreement with the measurements.
Before the cut-off frequency of the SHU medium, $f_\mathrm{c}$, only forward scattering occurs which leads to the cancellation of the effective attenuation. Nevertheless wave scattering causes wave dispersion inside the medium.
A main result was to show that transparency remains for a relatively high density medium, which confirms the suggestion of Leseur \textit{et al.} \cite{leseur2016high}. A band gap due to destructive interference originated from high position correlations between scatterers has been evidenced.
More importantly, we also found that transparency and band gap are invariant to translation along the medium.

By conducting measurements of the directivity of different SHU distributions, we were able to extend the comprehension of SHU properties by characterizing the Green's function.
If SHU media have a stealthiness degree such that no principal directions are found in the reciprocal space, the transparency is isotropic as well as the complete and wide band gaps that appear in transmission.

The results obtained should contribute to a better understanding of the scattering properties of SHU media. They also should have important applications in the context of designing heterogeneous media to control of wave propagation, or frequency filters.

\end{document}